\documentclass[
aps,
pre,
amsmath,amssymb,
reprint,
showkeys
]{revtex4-2}

\usepackage[T1]{fontenc}
\usepackage[utf8]{inputenc}
\usepackage{amsmath}
\usepackage{amssymb}
\usepackage{graphicx}
\usepackage{refstyle}
\usepackage{multirow}
\usepackage[
citecolor=blue,
colorlinks,
linkcolor=blue,
urlcolor=blue,
]{hyperref}
\usepackage[dvipsnames]{xcolor}
\usepackage{dcolumn}
\usepackage{bm}
\usepackage{soul}
\usepackage[T1]{fontenc}
\usepackage{xcolor}
\usepackage{bm}
\usepackage[version=4]{mhchem}
\usepackage{tikz}
\usetikzlibrary{arrows,shapes,trees}

\begin{document}
	\title{Eco-evolutionary games for harvesting self-renewing common resource: Effect of growing harvester population}
\author{Joy Das Bairagya}
\email{joydas@iitk.ac.in}
\affiliation{
	Department of Physics,
	Indian Institute of Technology Kanpur,
	Uttar Pradesh 208016, India
}
\author{Samrat Sohel Mondal}
\email{samrat@iitk.ac.in (corresponding author)}
\affiliation{
	Department of Physics,
	Indian Institute of Technology Kanpur,
	Uttar Pradesh 208016, India
}
\author{Debashish Chowdhury}
\email{debch@iitk.ac.in}
\affiliation{
	Department of Physics,
	Indian Institute of Technology Kanpur,
	Uttar Pradesh 208016, India
}
\author{Sagar Chakraborty}
\email{sagarc@iitk.ac.in}
\affiliation{
	Department of Physics,
	Indian Institute of Technology Kanpur,
	Uttar Pradesh 208016, India
}
\date{\today}

\begin{abstract}
	The tragedy of the commons (TOC)  is a ubiquitous social dilemma witnessed in interactions between a population of living entities and shared resources available to them: The individuals in the population tend to selfishly overexploit a common resource as it is arguably the rational choice, or in case of non-human beings, it may be an evolutionarily uninvadable action. How to avert the TOC is a significant problem related to the conservation of resources. It is not hard to envisage situations where the resource could be self-renewing and the size of the population may be dependent on the state of the resource through the fractions of the population employing different exploitation rates. If the self-renewal rate of the resource lies between the maximum and the minimum exploitation rates, it is not a priori obvious under what conditions the TOC can be averted. In this paper, we address this question analytically and numerically using the setup of an evolutionary game theoretical replicator equation that models the Darwinian tenet of natural selection. Through the replicator equation, while we investigate how a population of replicators exploit the shared resource, the latter's dynamical feedback on the former is also not ignored. We also present a transparent bottom-up derivation of the game-resource feedback model to facilitate future studies on the stochastic effects on the findings presented herein.
\end{abstract}
\maketitle
\section{Introduction}
\begin{figure*}
	\centering
	\includegraphics[width=1.0\linewidth]{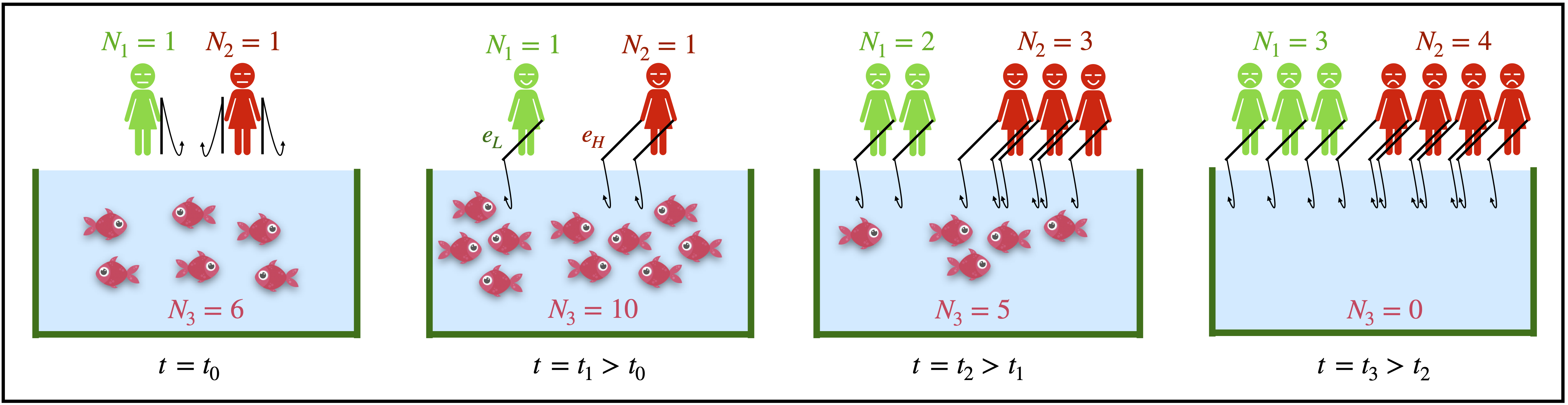}
	\caption{Illustrating game-environment feedback and the tragedy of the commons: $N_1$ number of type A individuals (`green' consumers with harvesting rate $e_L$) and $N_2$ number of type B individuals (`red' consumers with harvesting rate $e_H>e_L$) harvest $N_3$ unit of self renewing common resource (whose growth rate is $r_b$). Due to the higher harvest rate, the frequency of the type B individuals increases more than that of the type A. Although the resource keeps trying to renew itself, its overexploitation may eventually (here at time $t=t_3$) leads to the collapsing tragedy of the commons, i.e., $N_3=0$.}
	\label{fig:fig1}
\end{figure*}
Survival of an individual's lineage amid threats from the opponents in a population is dependent on its fitness that is highly influenced by the surrounding environment to which the population adapts. A favourable environmental state is beneficial for the individual and, partially, it is in the hands of the individual to keep the environment favourable. Unfortunately, the myopic selfishness of individuals involved in strategic interactions with the rest of the population may lead to overexploitation of the resources present in the environment and, thus, the environmental state is degraded to such a level that the fitness takes a hit. This degradation of the environment is commonly known as the tragedy of the commons (TOC)~\cite{lloyd1833two, hardin1968s, ostrom1999coping}. The population under question need not be human; it could even be non-human population (e.g, that of plants~\cite{Zhang2000,gersani2001tragedy,Falster2003,Day2003,zea2006tragedy}, animals~\cite{Dall2005,Rankin2006}, insects~\cite{Frank1995,Fournier2003,Wenseleers2004,Helanter2004}, microbes~\cite{Frank1996,Rainey2003,Hodgson2004,Roode2005,MacLean2006, Kerr2006}, cancer cells~\cite{Steven2004}, etc.) in which the individuals' cognitive capacity and rationality are almost next to none when compared with that of the modern humans.

In the context of the TOC in general biological evolutionary systems, the commons, i.e., the resources, are categorized into three types~\cite{Rankin2007}: extrinsic goods, social goods created by mutual cooperation, and social goods formed due to restraining from conflicts. Moreover, whether the TOC means complete or partial devastation of the resource leads to the naming of the respective tragedy as collapsing or component. This paper looks to investigate both collapsing and component tragedies of extrinsic goods exploited by the individuals evolving in line with the theory of natural selection~\cite{fisher1930book} and their feedback on the fitnesses of the individuals.

To this end, recently~\cite{weitz2016pnas,tilman2020nc} the formalism of evolutionary game theory~\cite{smith1972book,smith1973nature,smith1974jtb,smith1982book} and the deterministic replicator dynamics~\cite{taylor1978mb,schuster1983jtb,schuster1985bbpc,cressman2014pnas} have proven to be very insightful in studying the dynamics of the aforementioned feedback. However, the motivation behind all the governing equations therein is rather phenomenological. In the light of the existing literature on microscopic stochastic birth-death process~\cite{traulsen2005prl,lin2019prl,2021MCChaos} leading to replicator equation in the mean field limit and that on the microscopic process leading similarly to deterministic population dynamics equation (e.g., logistic equation~\cite{mendez2015stochastic}), in this paper we are motivated to derive the macroscopic model of the game-environment feedback dynamics starting from a microscopic multidimensional birth-death process that incorporates the feedback between the evolutionary system and an ecological self-renewing resource. Furthermore, this endeavour automatically facilitates the study of finite sized population with a finite carrying capacity which is known to have quite non-trivial counter-intuitive effects~\cite{Bairagya2021}. In passing, we remark that there do exist extensive investigations using stochastic models~\cite{Melbinger2010,Cremer2011,Melbinger2015,Wienand2017,Wienand2018} that, however, do not include an explicit equation for resource's dynamics and capture the environmental changes mostly through rather ad-hoc time-dependent carrying capacity.

After we derive the macroscopic deterministic eco-evolutionary dynamics, we focus on the surprisingly overlooked~\cite{tilman2020nc} scenario where the intrinsic growth rate of the resource is intermediate between the rates with which it is harvested by the population of selfish individuals. Such a setup is naturally suited to address both the collapsing and the component versions of the TOC; for the growth rate being lower than the lowest harvesting rate, only the collapsing TOC is expected, while for the growth rate being higher than the highest harvesting rate, only the component TOC is expected. We elaborately discuss the possible dynamics outcomes and also investigate the potential role of the finite carrying capacity of the population in averting the TOC. However, before we discuss these results below, we systematically explain and derive the mathematical model on which our entire paper is based.

\section{Deriving the model}	

Let us assume that the individuals (players) of a population harvest a shared resource pool whose instantaneous state is denoted by denoted $n(t)$ that is normalized such that $0 \leq n(t) \leq 1$. Let $N_i(t)$ be the number of the individuals who are using $i$-th  action and $N(t) = \sum_{i} N_{i}(t)$ be instantaneous size of the population. Obviously, $x_i(t)=N_i(t)/N$ is the fraction of the total population using $i$th action at any instant of time $t$. The state of the harvesting population can be written as a frequency vector ${\bm x}(t) \equiv [x_1(t),\, x_2(t),\, \cdots,\, x_{\mu}(t)]$, each element of which is also interpretable as the probabilities of choosing $i$-th action by an arbitrary agent from that population. 

As far as  the evolution of ${\bm x}(t)$ is concerned, within the paradigm of natural selection, the standard replicator equation is the most appropriate model. However, in this paper, we want to see the effect of finite carrying capacity in the coupled dynamics of the replicator and the resource by allowing feedback between the resource and the players. Therefore, we intend to extend the replicator equation so as to derive the set of equations that describes the time evolution of the composite state, $\sigma(t) \equiv  [n(t), {\bm x}(t), N(t)]$, of the full eco-evolutionary system described above. We would like to derive the equations starting from the microscopic birth-death process. In the simplest non-trivial setting, we choose $\mu=2$ and therefore the composite state can be alternatively and equivalently represented by ${\bm N}\equiv[N_1(t),\,N_2(t),\,N_3(t)]$ where $N_3(t)$ is unnormalised $n(t)$. The motivation behind this choice is made clear in the following discussion.

\subsection{Stochastic birth-death process}
For the sake of concreteness, without any loss of generality of the question we are interested in, consider the following scenario (see FIG.~\ref{fig:fig1}): In a consumer population of size $N$, $N_1$ number of (type A) players harvest a resource with a rate $e_L$ that is lower than the harvesting rate $e_H$ employed by the rest $N_2=N-N_1$ number of players (type B). The resource's state is characterized by $N_3$ harvestable units of C (say); e.g., fishes in a pond, trees in a forest, bacteria targeted by bacteriophages, etc. The resource is self-renewing, meaning it has a birth rate ($r_b$). Additionally, due to competition, there is a death rate ($r_d$). The effective birth rate of a harvester depends on the common resource ($N_3$) and the type of opponent it interacts with. There are death rates due to interactions between the harvesters; the death rates are assumed to be independent of the resource state that only assists in contributing to the fitness of the harvesters. The entire birth-death process can, thus, be represented as follows:
\begin{center}
    \ce{A + A  ->[$b_{11}(N_3)$] A + A + A}\\
	\ce{B + B  ->[$b_{22}(N_3)$] B + B + B}\\

    \begin{tikzpicture}
		\node (a) at (0, 0) {A+B}; 
		\node (b) at (2.5, 0.3){A+B+A};
		\node (c) at (2.5, -0.3){A+B+B};
		\draw (a) -- (0.7,0)--(0.7,0.3);
		\draw[->](0.7,0.3)-- (b) node[pos=0.5,above] {$\scriptstyle{b_{12}(N_3)}$};
		\draw (a)-- (0.7,0)--(0.7,-0.3);
		\draw[->](0.7,-0.3)-- (c) node[pos=0.5,above] {$\scriptstyle{b_{21}(N_3)}$};
	\end{tikzpicture}

	\ce{A + A ->[$d'_{11}$] A}\\
	\ce{B + B ->[$d'_{22}$] B}\\

    \begin{tikzpicture}
		\node (a) at (0, 0) {A+B}; 
		\node (b) at (1.5, 0.3){B};
		\node (c) at (1.5, -0.3){A};
		\draw (a) -- (0.7,0)--(0.7,0.3);
		\draw[->](0.7,0.3)-- (b) node[pos=0.5,above] {$\scriptstyle{d'_{12}}$};
		\draw (a)-- (0.7,0)--(0.7,-0.3);
		\draw[->](0.7,-0.3)-- (c) node[pos=0.5,above] {$\scriptstyle{d'_{21}}$};
	\end{tikzpicture}

	\ce{C ->[$r_b$] C + C}\\
	\ce{C + C ->[$r_d$] C}\\
	\ce{C  ->[$xe_L + (1-x)e_H$] $\phi$}
\end{center}

Here $b_{ij}$'s are the birth rates that depend on the resource state $N_3$ and $d'_{ij}$'s are the death rates. Also, here $x\equiv N_1/N$ and hence $1-x=N_2/N$ are the instantaneous fractions of type A and type B harvesters, respectively; therefore, there is depletion of the resource by the harvesters at a rate $xe_L + (1-x)e_H$. 

Proceeding further, we can now calculate the transition probabilities per unit time. The transition  rates of type A or type B individuals to increase by one are respectively given by
\begin{equation}
T^+_1(\bm{N})={N}_1 \left[b_{11}(N_3)\frac{{N}_1-1}{{N} -1} +  b_{12}(N_3)\frac{{N}_2}{{N} -1}\right],
\end{equation}
and
\begin{equation}
T^+_2(\bm{N})={N}_2 \left[b_{21}(N_3)\frac{{N}_1}{{N} -1} +  b_{22}(N_3)\frac{{N}_2-1}{{N} -1}\right].
\end{equation}
Similarly, the transition  rates for going from $({N}_1, {N}_2,N_3)$ to $({N}_1-1, {N}_2,N_3)$ and from $({N}_1, {N}_2,N_3)$ to $({N}_1, {N}_2-1,N_3)$ respectively are
\begin{equation}
T^-_1(\bm{N})={N}_1 \left(d'_{11}\frac{{N}_1-1}{{N} -1} +  d'_{12}\frac{{N}_2}{{N} -1}\right),
\end{equation}
and
\begin{equation}
T^-_2(\bm{N})={N}_2 \left(d'_{21}\frac{{N}_1}{{N} -1} +  d'_{22}\frac{{N}_2-1}{{N} -1}\right).
\end{equation}
Furthermore, the transition rate of going from $N_3$ units of self-renewing common resource to $(N_3+1)$ units is for any state is
\begin{equation}
T_{3}^+(\bm{N})= N_3 r_b.
\end{equation}
Lastly, the probability of going from $N_3$ to $N_3-1$ per unit time is 
\begin{align}
T_{3}^-(\bm{N})=\frac{N_3(N_3-1)}{2}r_d + N_3 \left(xe_L+(1-x)e_H\right),
\end{align}
where the first term on the right hand side accounts for the number of possible matched pairs leading to the possible decline in the resource, and the second term accounts for the decline due to harvesting.

Let $P(N_1,N_2,N_3,t)$ be the probability of having ${N}_1$ number of type A and ${N}_2$ number of type B individuals harvesting $N_3$ units resource at an instant of time $t$. Suppose that the carrying capacities of  type A and type B consumers, and the resource are $K_1$, $K_2$ and $K_3$ respectively. One should choose $K_1=K_2$ as the carrying capacity of the consumer population. Thus, we can normalize the variables $N_1,N_2, N_3$ as follows: $\tilde{N_i}=N_i/K_i$. The probability density function in terms of these normalized coordinates is $\rho(\tilde{N_1},\tilde{N_2},\tilde{N_2},t)=P(N_1,N_2,N_3,t)K_1K_2K_3$. The discrete master equation for the model can therefore be written as,
\begin{align}
	&\rho(\tilde{N_1},\tilde{N_2},\tilde{N_2},,t+\delta t)=\nonumber\\
	&\sum_{\alpha_1,\alpha_2,\alpha_3}\rho(\tilde{N_1}+\frac{\alpha_1}{K_1},\tilde{N_2}+\frac{\alpha_2}{K_2},\tilde{N_3}+\frac{\alpha_3}{K_3},t)\nonumber\\
	&\phantom{\sum_{\frac{\alpha_1}{K},\frac{\alpha_2}{K},\frac{\alpha_3}{k}}}\times\Gamma^{-\alpha_1}_1(\tilde{N_1}+\frac{\alpha_1}{K_1},\tilde{N_2}+\frac{\alpha_2}{K_2},\tilde{N_3}+\frac{\alpha_3}{K_3},t)\nonumber\\
	&\phantom{\sum_{\frac{\alpha_1}{K},\frac{\alpha_2}{K},\frac{\alpha_3}{k}}}\times\Gamma^{-\alpha_2}_2(\tilde{N_1}+\frac{\alpha_1}{K_1},\tilde{N_2}+\frac{\alpha_2}{K_2},\tilde{N_3}+\frac{\alpha_3}{K_3},t)\nonumber\\
	&\phantom{\sum_{\frac{\alpha_1}{K},\frac{\alpha_2}{K},\frac{\alpha_3}{k}}}\times\Gamma^{-\alpha_3}_3(\tilde{N_1}+\frac{\alpha_1}{K_1},\tilde{N_2}+\frac{\alpha_2}{K_2},\tilde{N_3}+\frac{\alpha_3}{K_3},t),
	\label{MasterA}
	\end{align}
	where $\alpha_i \in \{-1,0,1\}$, $\Gamma^{-1}_i(\tilde{\bm{N}},t)\equiv T^-_i(\tilde{\bm{N}},t)\delta t$, 
  $\Gamma^{0}_i(\tilde{\bm{N}},t)\equiv(1-T^+_i(\tilde{\bm{N}},t)\delta t-T^-_i(\tilde{\bm{N}},t)\delta t)$, and $\Gamma^{1}_i(\tilde{\bm{N}},t)\equiv T^+_i(\tilde{\bm{N}},t)\delta t$ $\forall i\in\{1,2,3\}$. 
 In the continuum limit, i.e., $1/K_i \to 0$ and $\delta t\to 0$, we can write the corresponding Fokker--Planck equation as follows:
\begin{widetext}
\begin{align}
\frac{\partial{\rho(\tilde{\bm{N}},t)}}{\partial{t}}=&- \sum_{i={1,2,3}} \frac{1}{K_i} \frac{\partial}{\partial{{\tilde{N}}_{i}}}\left[(T^+_{i}(\tilde{\bm{N}})-T^-_{i}(\tilde{\bm{N}}))\rho(\tilde{\bm{N}},t)\right]+\frac{1}{2}\sum_{i={1,2,3}} \frac{1}{K^2_i}  \frac{\partial^2}{\partial{{\tilde{N}}_{i}^2}}\left[(T^+_{i}(\tilde{\bm{N}})+T^-_{i}(\tilde{\bm{N}}))\rho(\tilde{\bm{N}},t)\right]\nonumber\\&+\frac{1}{2}\sum_{i={1,2,3}} \frac{1}{K_i K_j} \frac{\partial^2}{\partial{{\tilde{N}}_i}\partial{{\tilde{N}}_j}}\left[\left(T^+_{i}(\tilde{\bm{N}})-T^-_{i}(\tilde{\bm{N}})\right)\left(T^+_{j}(\tilde{\bm{N}})-T^-_{j}(\tilde{\bm{N}})\right)\rho(\tilde{\bm{N}},t)\right]\left( 1-\delta_{ij}\right).
\label{FK}
\end{align}
\end{widetext}

\subsection{Mean-field equations}
From Eq.~(\ref{FK}), using standard arguments~\cite{Gillespie1996,Vankampen}, we can write the mean field dynamics of 
$\overline{N}_i\equiv K_i \int \tilde{N}_i \rho(\tilde{\bm{N}},t) d \tilde{\bm {N}} $ as
\begin{equation}
\frac{d\overline{{N}}_i}{dt}= T^+_{i}(\overline{\bm{N}})-T^-_{i}(\overline{\bm{N}}).
\label{n_i dot}
\end{equation}
Therefore, explicitly we can write the dynamics in terms of the mean total number of the individuals (${\overline{{N}}}\equiv  \overline{N}_1 +\overline{N}_2$), the mean fraction ($\overline{x}\equiv\overline{N}_1/N$) of type A individuals, and the normalization mean number of resource population ($\overline{n}\equiv\overline{N}_3/\mathcal{N}$; $\mathcal{N}$ is some normalization constant that we shall elaborate on later in the paper)  as respectively written below:
\begin{subequations}
\begin{eqnarray}
\frac{d\overline{ N}}{dt}&=&{\overline{N}} \overline{\pi}(x,\overline{N},\overline{n})\label{Ndot},\qquad\qquad\\
\frac{d{\overline x}}{dt}&=&{\overline x}(1-{\overline x})\left[\pi_1({\overline x},\overline{N},\overline{n})-\pi_2({\overline x},\overline{N},\overline{n})\right] \label{xdot},\qquad\qquad\\
\frac{d\overline{n}}{dt}&=& r_b\overline{n}\left(1-\frac{\overline{n}}{k/\mathcal{N}}\right)-\overline{n}\left[\overline{x}e_L+(1-\overline{x})e_H\right].\label{N3dot}\qquad
\end{eqnarray}
\end{subequations}
Here, with $d_{ij}\equiv d'_{ij}/N~\forall i,j\in \{1,2\}$, we have defined,
\begin{subequations}
\begin{eqnarray}
\pi_1(\overline{x},\overline{N},\overline{n})&\equiv&b_{11}(\overline{n})\overline{x}\left[1-\frac{\overline{N}}{\frac{b_{11}(\overline{n})}{d_{11}}}\right] \nonumber\\
&&+b_{12}(\overline{n})(1-\overline{x})\left[1-\frac{\overline{N}}{\frac{b_{12}(\overline{n})}{d_{12}}}\right], \\
\pi_2(\overline{x},\overline{N},\overline{n})&\equiv&b_{21}(\overline{n})\overline{x}\left[1-\frac{\overline{N}}{\frac{b_{21}(\overline{n})}{d_{21}}}\right]\nonumber\\
&&+b_{22}(\overline{n})(1-\overline{x})\left[1-\frac{\overline{N}}{\frac{b_{22}(\overline{n})}{d_{22}}}\right], \\
\overline{\pi}(\overline{x},\overline{N},\overline{n})&\equiv& \overline{x}\pi_1(\overline{x},\overline{N},\overline{n})+(1-\overline{x})\pi_2(\overline{x},\overline{N},\overline{n}),\label{r_N(x)}\qquad\qquad\\
k&\equiv&K_3=\frac{r_b}{2r_d}.
\end{eqnarray}
\end{subequations}
Having obtained the mean field equations [Eq.~(\ref{Ndot})--(\ref{N3dot})] that describes the composite deterministic state $\bm{\sigma}(t)$, we henceforth drop the overhead bar for denoting the mean quantities for the sake of notational convenience without any scope of ambiguity.

The birth rates are assumed to be modulated by the resource availability; there is an implicit ecological resource (with carrying capacity in general, dependent on $n$ and interaction between harvesters) facilitating the growth of the population. The harvesting of the common resource leads to further enhancement of the birth rates that, however, may be different for different states of the resource. The simplest forms of the birth rates could be
\begin{subequations}
\begin{eqnarray}
&b_{11}({n})=(1-{n})R_0+ {n} R_1,\\
&b_{12}({n})=(1-{n})S_0+ {n} S_1, \label{coupling}\\
&b_{21}({n})=(1-{n})T_0 + {n} T_1,\\
&b_{22}({n})=(1-{n})P_0+ {n} P_1. 
\end{eqnarray}
\end{subequations}
Here, $R_0,~R_1,~S_0,~S_1,~T_0,~T_1,~P_0,$ and $P_1$ are nonnegative numbers. This immediately furnishes a game theoretic viewpoint of the situation at hand: Any arbitrarily chosen individual can have two strategies---harvesting with low rate $e_L$ and harvesting with high rate $e_H>e_L$---and the corresponding payoff matrix can be expressed as

\begin{eqnarray}
{\sf{U}}&&(n,N)=n\left[\begin{matrix}
R_1\left(1-\frac{N}{K_{11}(n)}\right)  &  S_1 \left(1-\frac{N}{K_{12}(n)}\right)\\
\\
T_1 \left(1-\frac{N}{K_{21}(n)}\right) &  P_1\left(1-\frac{N}{K_{22}(n)}\right)\\
\end{matrix}\right]\nonumber\quad\\
&&\quad+(1-n)\left[\begin{matrix}
R_0\left(1-\frac{N}{K_{11}(n)}\right)  &  S_0 \left(1-\frac{N}{K_{12}(n)}\right)\\
\\
T_0 \left(1-\frac{N}{K_{21}(n)}\right) &  P_0\left(1-\frac{N}{K_{22}(n)}\right)\\
\end{matrix}\right],\qquad
\label{pay-off matrix}
\end{eqnarray}
where $K_{ij}(n)\equiv b_{ij}(n)/d_{ij}$ ~$\forall i,j\in\{1,2\}$ are the carrying capacities. Therefore, the $i$th type (either type A or type B) individual's fitness, which depends on the state of the composite system and frequency of the types, is $f_{i}(\boldsymbol\sigma) = \sum_{j=1}^{2} {\sf U}_{ij}({\bm n},N) x_{j}.
$ Hence, the final eco-evolutionary dynamics of the composite system, that we were seeking in the beginning of this section, can be cast in the following form [see Eq.~(\ref{Ndot})--(\ref{N3dot})]:
\begin{subequations}
\begin{eqnarray}
	\frac{dN}{dt} &=& N\sum_{i=1}^{2} \sum_{j=1}^{2} {\sf U}_{ij}(n,N) x_{i}x_j,
	\label{N_eq_model}\\
\frac{dx}{dt} &=&  x\left[\sum_{j=1}^{2} {\sf U}_{1j}(n,N) x_{j} - \sum_{i=1}^{2}\sum_{j=1}^{2} {\sf U}_{ij}(n,N)  x_ix_{j}\right],\qquad\,\label{x_eq_model}\\
\frac{d{n}}{dt}&=& r_b{n}\left(1-\frac{{n}}{k/\mathcal{N}}\right)-{n}\left[{x}e_L+(1-{x})e_H\right].\label{n_eq_model}\qquad
\end{eqnarray}
\end{subequations}
\subsection{The relevant resource dynamic}
\label{parameters}
The fate of the resource, with a given carrying capacity $k$, is explicitly determined by three factors---its intrinsic growth rate and the rates at which it is harvested. Since by definition, $e_H>e_L$, following three scenarios  are exhaustive:
\begin{enumerate}
\item {$r_b<e_L<e_H$}:
This case is rather trivial as the intrinsic growth rate of the self-renewing resource is unable to sustain the resource against its constant depletion at relatively higher rates. Note that $r_b[1-n/(k/\mathcal{N})]<xe_L+(1-x)e_H$ since $x,n/k\in[0,1]$. Hence, eventually $n=0$ because $dn/dt<0$ $\forall x,n$; i.e., TOC is inevitable for such a resource. We are not concerned with this trivial case in this paper.
\item{$r_b>e_H>e_L$}: In the other extreme, if the intrinsic growth rate is higher than the high harvest rate, it is easily seen that about $n=0$ state $dn/dt>0$ leading to unconditional prevention of the collapsing TOC. Of course, the resource state cannot reach its maximum potential, $n=k$, in the presence of the harvesters. In other words, the component TOC is may still realized. To mathematically capture this scenario, it was shown~\cite{tilman2020nc} that it is convenient to choose $\mathcal{N}=(e_H-e_L)k/r_b$. Furthermore, $n=(r_b-e_H)/(e_H-e_L)$ could be choosen as the new origin so that the rescaled $n$ (at all times) is normalized to remain between zero and one. The negative (rescaled) resource state is never accessible by the construction of the model; for the same reason, it does not make sense to take any initial condition negative. Note, however, that shifting the origin to $-(r_b-e_H)/(e_H-e_L)$ is merely a convention suited to the scenario under consideration. In general, this shift is unnecessary, as is crystal clear in the next case.
\item{$e_H>r_b>e_L$}: This is arguably the most interesting case because there is a trade-off  between the low and the high harvesting rates for the resource with an intrinsic growth rate lying between the two harvesting rates. Here, collapsing TOC is present as well. Hence, it is natural to choose $\mathcal{N}=k(1-e_L/r_b)$---the maximum achievable value of the resource---and recast Eq.~(\ref{N3dot}) for this case as
\begin{equation}
\qquad\,\,\frac{1}{\varepsilon}\frac{d{n}}{dt}=r_b{n}\left[1-\left(1-\frac{e_L}{r_b}\right){n}\right]-{n}\left[{x}e_L+(1-{x})e_H\right].\label{ndot}
\end{equation}
Here, we have included a factor $\varepsilon<1$ that models the fact that the environment commonly evolves at a slower timescale than the replicators; mathematically, it merely amounts to redefining $r_b$, $e_H$, and $e_L$. Under the present normalization, the minimum resource state (that was made the origin in case 2 above) when $r_b>e_H>e_L$ is $n=(r_b-e_H)/(r_b-e_L)$ which corresponds to the component TOC. The minimum resource state, when {$e_H>r_b>e_L$}, is simply zero, which corresponds to the collapsing TOC.
\end{enumerate}
The rest of the paper deals with the eco-evolutionary game-environment feedback dynamics governed by Eq.~(\ref{N_eq_model}), Eq.~(\ref{x_eq_model}), and Eq.~(\ref{ndot}). We remark that Eq.~(\ref{ndot}) is not exclusive to case 3 (i.e., {$e_H>r_b>e_L$}) and is of most general validity; however, only for this case, the non-negativity of $n$ at all times is guaranteed. 

\section{Results}
We now plan to carry out stability analysis of various possible eventualities of the composite system, and hence find out when TOC could be averted and also study the effect of the finiteness of the carrying capacities. However, first we must contrast the cases $r_b>e_H>e_L$ with $e_H>r_b>e_L$ in when all the carrying capacities $K_{ij}\to \infty~\forall i,j\in\{1,2\}$  that happens when $d_{ij} \to 0~\forall i,j\in\{1,2\}$. 
\subsection{Infinite carrying capacity}
%

\begin{figure*}
	\centering
	\includegraphics[width=1\linewidth]{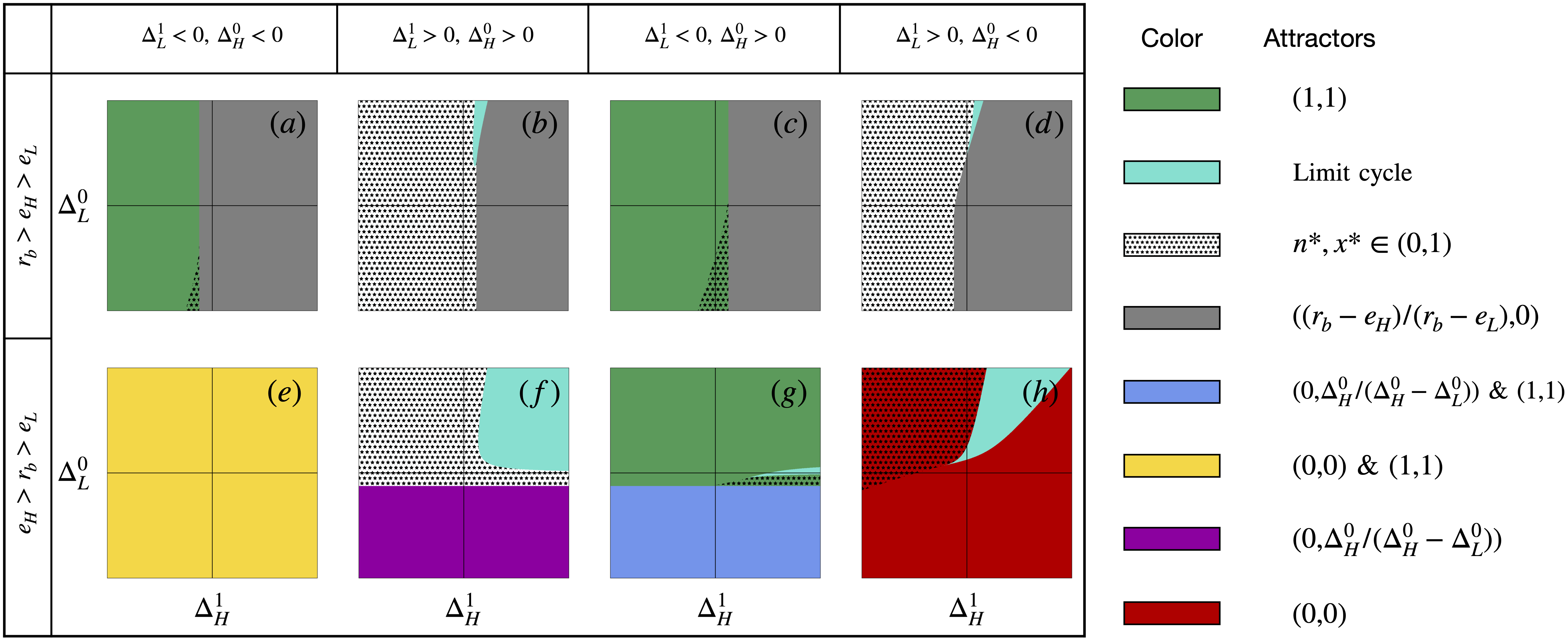}
	\caption{All possible distinct eventual outcomes for the composite state $(n,x)$ in the population with infinite carrying capacity: We vary both $\Delta^0_L$ and $\Delta^1_H$ from $-8$ to $+8$ while keeping  $\Delta^1_L$ and $\Delta^0_H$ fixed such that the four columns of subplots correspond to four exhaustive combinations of signs of $\Delta^1_L$ and $\Delta^0_H$. For the sake of concreteness and illustrative purpose, and without any loss of generality, we have fixed $\Delta^1_L$ and $\Delta^0_H$ to $-1$ and $-1$, $+1$ and $+1$, $-1$ and $+1$, and $+1$ and $-1$ respectively in the four consecutive columns.  The upper row, where $r_b>e_H>e_L$, exhibits realization of component TOC at worst; whereas the lower row, where $e_H>r_b>e_L$, depicts realization of collapsing TOC. Again, for concreteness and illustrative purpose, we have chosen $(r_b,e_H,e_L)$ to be $(1.5,1.0,0.5)$ in the former and $(1.0,1.5,0.5)$ in  the latter. Here $\varepsilon=0.2$; other values of $\varepsilon$ produce qualitatively equivalent figures.}
	\label{fig2_comparision}
\end{figure*}
%
In this limit, implicitly one is assuming that the consumer population has grown to its infinite capacity and is being kept fixed there. Therefore, the dynamics becomes two-dimensional because only Eq.~(\ref{x_eq_model}), and Eq.~(\ref{ndot}) are relevant. Moreover, the payoff matrix given in Eq.~(\ref{pay-off matrix}) is also simplified to
\begin{equation}
{\sf{U}}({n})={n}\left[\begin{matrix}
R_1  &  S_1 \\
T_1 &  P_1\\
\end{matrix}\right]+(1-{n})\left[\begin{matrix}
R_0  &  S_0 \\
T_0  &  P_0\\
\end{matrix}\right].
\label{pay-off matrix for infinite carrying capacity}
\end{equation}

\subsubsection{Linear stability results}
The fate of the resource and the fractions of the harvester types depend on the intrinsic growth rate of the resource and the harvesting rates, and on the payoffs. Specifically, the dependence on payoff enters through the idea of incentives: In an interaction of two individuals, incentive measures the change in the payoff of the focal player on unilaterally shifting from her current strategy to the other one. We define the following four incentives:
\begin{enumerate}
	\item  $\Delta_H^{0}\equiv S_0 - P_0$ is the incentive for low harvesting in the depleted common resource state, i.e., $n=0$, when in an interaction with a high harvester;
	\item $\Delta_L^{0}\equiv R_0 - T_0$ is the incentive for low harvesting in the depleted common resource state, i.e., $n=0$, when in an interaction with a low harvester.
	\item $\Delta_H^{1}\equiv P_1 - S_1$ is the incentive for high harvesting in the replete common resource, i.e., $n=1$, when in an interaction with a high harvester;
	\item $\Delta_L^{1}\equiv T_1 - R_1$ is the incentive for high harvesting in the replete common resource state, i.e., $n=1$, when in an interaction with a low harvester;
\end{enumerate}

In the specific parameter range of our interest, i.e., $r_b>e_L$, the fixed points, $(n^*,x^*)$, can be put in one of the following three classes: fixed points associated with TOC $(n^*=0)$, fixed points associated with complete prevention of TOC $(n^*=1)$, and the ones associated with partial prevention of TOC $(0<n^*<1)$. These are written below:
\begin{enumerate}
	\item {Realization of TOC:} There are three fixed points corresponding to this case: $(0,0)$, $(0,1)$ and  $(0,\Delta^0_H/(\Delta^0_H-\Delta^0_L))$ . The second one is always unstable. The first fixed point is stable when inequalities $r_b<e_H$ and $\Delta^0_H<0$ hold simultaneously. Whereas the third fixed point is stable when $[(r_b-e_L)\Delta^0_H-(r_b-e_H)\Delta^0_L]/(\Delta^0_H-\Delta^0_L) <0$ and $\Delta^0_H \Delta^0_L/(\Delta^0_H-\Delta^0_L)<0$ hold simultaneously.
	\item{Complete prevention of TOC:} Only one fixed point $(1,1)$ corresponds to the complete prevention of TOC. This fixed point becomes stable only when $\Delta^1_L<0$.
	\item{Partial prevention of TOC:} This case has two fixed points associated with it. The first one $(n^*,x^*)=((r_b-e_H)/(r_b-e_L),0)$, which exists when $r_b>e_H$ and is stable if additionally $\left[ (e_H-e_L)\Delta^0_H-(r_b-e_H)\Delta^1_H\right] <0$. (This fixed point corresponding to the component TOC was reported as the TOC in the literature~\cite{tilman2020nc}). Less restrictive in existence is another fixed point, for which $0<x^*<1$ in addition to $0<n^*<1$, that exists whenever $r_b>e_L$. Analytically writing down its explicit expression and the condition for its stability is not possible. We have checked its existence and the condition for stability for a wide range of parameter values numerically that we have presented in FIG.~\ref{fig2_comparision}. More interestingly, this internal fixed point can give rise to a limit cycle attractor through the Hopf bifurcation with the change of parameters.
	\end{enumerate}

\subsubsection{Numerical results}
We use the hitherto gathered information to do appropriate numerics for two cases of $r_b>e_L$, viz., $r_b>e_H>e_L$ and $e_H>r_b>e_L$. First we reproduce the results~\cite{tilman2020nc} for the case $r_b>e_H>e_L$ (first row of FIG.~\ref{fig2_comparision}), however, now allowing for the initial states of the resource less than $(r_b-e_H)/(e_H-e_L)$. Subsequently, we present the new results concerning the case $e_H>r_b>e_L$ (second row of FIG.~\ref{fig2_comparision}) and compare with the former case.

What we plot in FIG.~\ref{fig2_comparision} are the all possible eventualities for the resource state and the cooperator fraction starting from arbitrary initial conditions. Mathematically, all possible eventualities are captured by the attractors, viz., stable fixed points and stable limit cycle, of the system. Obviously, the existence of these attractors depends on the specific combination of values of the four incentives. A few case studies on the real-life applications of some of the combination is presented in a recent paper~\cite{tilman2020nc}; more can be contemplated in various realistic situations. However, in this paper, our motivation is to be theoretically exhaustive rather than being focussed on discussing some specific cases.

The central difference between the two aforementioned cases is that for $r_b>e_H>e_L$, there is no collapsing TOC, unlike what is realized for $e_H>r_b>e_L$. In the former, the state of the resource lies between $(r_b-e_H)/(e_H-e_L)$ to $1$, although the fraction of the low harvesters can go to zero. Even in the exclusive presence of the high harvesters, the resource is not fully exhausted (the grey region in FIG.~\ref{fig2_comparision}(a)--(d)); this is simply because the resource's self-renewal rate is higher than the high harvesting rate. Another interesting observation is that the limit cycle attractors appear only when $\Delta_L^1>0$ in the former case, whereas in the later case, this oscillatory partial prevention of the TOC is possible even when $\Delta_L^1<0$ (the cyan region in FIG.~\ref{fig2_comparision}(g)).

In the case $r_b>e_H>e_L$, there is bistability for $\Delta_L^1<0$ (FIG.~\ref{fig2_comparision}(a) and (c)): the partial prevention of TOC coexists with complete prevention of the TOC. While such a bistability is present even in the case $e_H>r_b>e_L$ (FIG.~\ref{fig2_comparision}(g)), many more types of bistability shows up in this case: coexistence between complete TOC (with zero cooperators) and complete prevention of TOC~(FIG.~\ref{fig2_comparision}(e)), coexistence between complete TOC (with nonzero cooperators) and complete prevention of TOC~(blue region in FIG.~\ref{fig2_comparision}(g)), and coexistence between complete TOC (with zero cooperators) and partial prevention of TOC~(spotted red region in FIG.~\ref{fig2_comparision}(h)).

\subsection{Finite carrying capacity}
Having compared the case of $e_L<r_b<e_H$ with the case $r_b>e_H>e_L$, we are now well set to comprehend the effect of the finiteness of the carrying capacity in a growing population with focus on the former case which is richer in features. 
\subsubsection{Modified incentives}
In contrast to the infinite population case, here, the payoffs for choosing the low or high rates of harvesting at replete or deplete environment are not solely dependent on the birth rates; it also depends on the death rates. Naturally, the incentives defined earlier must be modified. We note that the incentives decide the stability of the fixed points, so we expect the modified incentives to do the same in the case of finite carrying capacity. Hence, we introduce the modified incentives whose utility will be apparent later. 

First consider $\tilde{\Delta}^0_L\equiv d_{21}(R_0/d_{11}-T_0/d_{21})=d_{21}[K_{21}(n=1)-K_{11}(n=1)]$. In words, $\tilde{\Delta}^0_L$ is the difference (scaled by $d_{21}$) between the carrying capacity at $n=0$ (hence the superscript $0$) of an individual choosing high harvesting rate and the carrying capacity of the individual alternatively choosing low harvesting rate while interacting with another individual choosing low harvesting strategy (hence the subscript $L$). If we adopt the convention that $d_{21}/d_{11}\to1$ as $d_{21},d_{11}\to0$, we note that $\tilde{\Delta}^0_L\to\Delta^0_L$ as the death rates vanish.

With similar interpretations, we furthermore define $\tilde{\Delta}^0_H$, $\tilde{\Delta}^1_L$, and $\tilde{\Delta}^1_H$ respectively as
\begin{eqnarray*}
&&d_{12}(S_0/d_{12}-P_0/d_{22})=d_{12}[K_{12}(n=0)-K_{22}(n=0)],\\
&&d_{21}(T_1/d_{21}-R_1/d_{11})=d_{21}[K_{21}(n=1)-K_{11}(n=1)],\\	
&&{\rm and}\\
&&d_{12}(P_1/d_{22}-S_1/d_{12})=d_{12}[K_{22}(n=1)-K_{12}(n=1)].
\end{eqnarray*}
They all become the usual incentives in the limit of death rates going to zero.

\begin{figure*}
	\centering
	\includegraphics[width=0.9\linewidth]{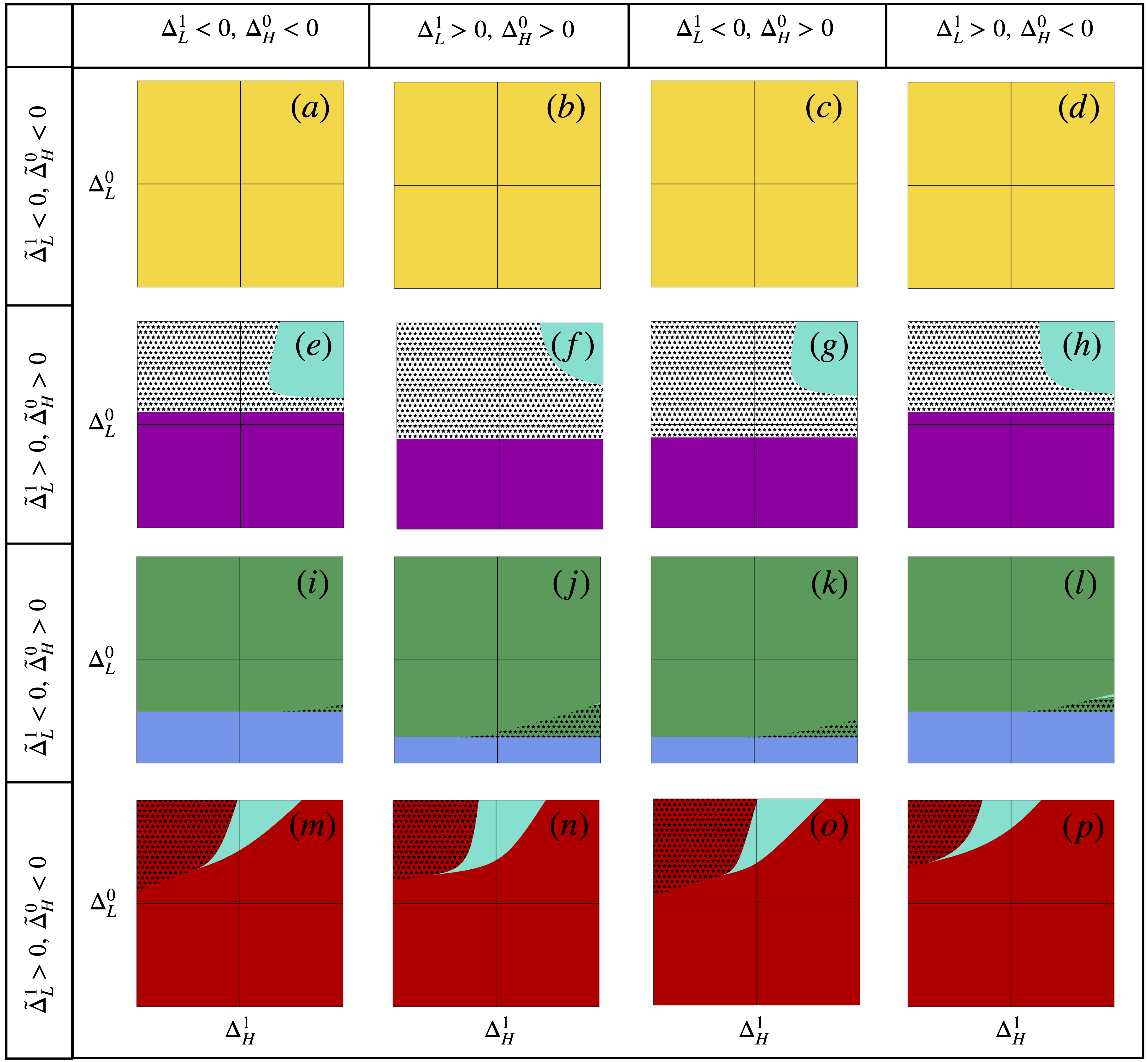}
	\caption{All eventual outcomes for the composite state $(n,x)$ in the population with finite carrying capacity and $e_H>r_b>e_H$: The colour code is same as used in FIG.~\ref{fig2_comparision}; additionally, in each case, $N^*$ takes some nonzero finite values which is not needed for our purpose. We vary both $\Delta^0_L$ and $\Delta^1_H$ from $-8$ to $+8$ while keeping  $\Delta^1_L$ and $\Delta^0_H$ fixed such that the four columns of subplots correspond to four exhaustive combinations of signs of $\Delta^1_L$ and $\Delta^0_H$. For the sake of concreteness and illustrative purpose, and without any loss of generality, we have fixed $\Delta^1_L$ and $\Delta^0_H$ to $-1$ and $-1$, $+1$ and $+1$, $-1$ and $+1$, and $+1$ and $-1$ respectively in the four consecutive columns. Similarly, we vary both $\tilde\Delta^0_L$ and $\tilde\Delta^1_H$ from $-8$ to $+8$ while keeping  $\tilde\Delta^1_L$ and $\tilde\Delta^0_H$ fixed such that the four columns of subplots correspond to four exhaustive combinations of signs of $\tilde\Delta^1_L$ and $\tilde\Delta^0_H$. In order to change the signs of these two modified incentives, we use four different combinations of death rates $d_{11}$, $d_{12}$, $d_{21}$, and $d_{22}$ (first row) $d_{11}=10^{-3}$, $d_{12}=1.5\times10^{-3}$, $d_{21}=1.5\times10^{-3}$, and $d_{22}=10^{-3}$; (second row) $d_{11}=1.5\times10^{-3}$, $d_{12}=10^{-3}$, $d_{21}=10^{-3}$, and $d_{22}=1.5\times10^{-3}$; (third row) $d_{11}=10^{-3}$, $d_{12}=10^{-3}$, $d_{21}=1.5\times10^{-3}$,  and $d_{22}=1.5\times10^{-3}$; (fourth row) $d_{11}=1.5\times10^{-3}$, $~d_{12}=1.5\times10^{-3}$, $d_{21}=10^{-3}$, and $d_{22}=10^{-3}$. We fix $R_1=S_1=R_0=S_0=5.0$. For concrete illustrative purpose, we have chosen $\varepsilon=0.2$; and $r_b$, $e_H$, and $e_L$ to be $1.0$, $1.5$, and $0.5$ respectively.}
	\label{fig3_effect_of_finiteness}
\end{figure*}	
\subsubsection{Linear stability results}

The first step of analysis is to study the features of the fixed point $(n^*,x^*,N^*)$ of the set of equations given by Eq.~(\ref{N_eq_model}), Eq.~(\ref{x_eq_model}), and Eq.~(\ref{ndot}). In the specific parameter range of our interest, i.e., $r_b>e_L$, there are some fixed points---viz, $(0,1,R_0/d_{11})$, $(0,0,0)$,  $(0,1,0)$, $(1,1,0)$, $((r_b-e_H)/(r_b-e_L),0,0)$, $(0,x^*\in[0,1],0)$ and $(n^*\in[0,1],x^*\in[0,1],0)$---that are always unstable and hence of no physical interest to us. As before, rest of the fixed points can be put in one of the following three exhaustive classes:
\begin{enumerate}
\item {Realization of TOC:}
The fixed point, $({n}^*,{x}^*,{N}^*)=(0,0,P_0/d_{22})$, corresponds to the TOC.  It is stable when $\tilde{\Delta}^0_H<0$ and when the higher harvesting rate is greater than the intrinsic growth of the resource, i.e., $e_H>r_b>e_L$. There is one more fixed point (with non-zero values of ${x^*}$ and ${N}^*$) which corresponds to TOC. However, its explicit closed analytical form is hard, and consequently, its stability analysis is analytically intractable as well. Hence, we are forced to undertake its analysis numerically to find when it exists and is stable, as is discussed later in this paper. 

\item {Complete prevention of TOC:}
The fixed point $({n}^*,{x}^*,{N}^*)=(1,1,R_1/d_{11})$, corresponds to the full prevention of the TOC. This fixed point becomes stable when the condition $\tilde{\Delta}^1_L<0$ is satisfied. 

\item {Partial prevention of TOC:}
This case has two fixed points. One is $({n}^*,{x}^*,{N}^*)=((r_b-e_H)/(r_b-e_L),0,[(e_H-e_L)P_0+(r_b-e_H)P_1]/[d_{22}(r_b-e_H)])$ that exists only when the intrinsic growth rate of the resource $r_b$ is higher than the higher harvesting rate $e_H$.  This prevention of the TOC when all the individuals are harvesting slower than the rate at which the self-renewing of the resource occurs is quite intuitive. This fixed point is stable when $(e_H-e_L)\tilde{\Delta}^0_H-(r_b-e_H)\tilde{\Delta}^1_H <0$. The other fixed point is the interior one, i.e., $x^*$ and $N^*$ are nonzero and non-unity. Again, its the analytical investigation is not conducive due to the absence of an explicit expression for it. This fixed point can undergo a change of stability through the Hopf bifurcation, and a limit cycle attractor can emerge; a stable limit cycle can also be interpreted to correspond to the partial prevention of TOC.
\end{enumerate}

\subsubsection{Numerical results}
Consider the fixed point $(0,0,P_0/d_{22})$ that corresponds to TOC. We have analytically seen that it is stable if the carrying capacity for the high rate harvesters is more than the low rate harvester in the depleted environment while interacting with another high rate harvester. This makes the low rate harvesters less effective, and eventually, all the individuals become high rate harvesters. Since the intrinsic growth rate of the resource is assumed to be less than the high harvesting rate, the TOC is inevitable in such a situation (FIG.~\ref{fig3_effect_of_finiteness}(a)--(d) and FIG.~\ref{fig3_effect_of_finiteness}(m)--(p)) as implied by the stability of the fixed point.

Similarly, the total prevention of TOC happens when the system converges to the fixed point $(1,1,R_1/d_{11})$. This fixed point becomes stable if the carrying capacity at $n=1$ for the low rate harvester is more than that of a high rate harvester while interacting with another low rate harvester. In that situation, the high rate harvesters become less effective, and all individuals eventually become low rate harvesters. Since the intrinsic growth rate of the resource is higher than the low harvesting rate, the TOC is completely averted in this case~(FIG.~\ref{fig3_effect_of_finiteness}(a)--(d) and FIG.~\ref{fig3_effect_of_finiteness}(i)--(l)). 

Other than these two fixed points, the stabilities of the other fixed points, and hence the corresponding state of the resource, depend on the modified incentives in nontrivial ways. So we study the system over a wide range of parameters space, and the exhaustive results are compactly presented in FIG.~\ref{fig3_effect_of_finiteness}.

The crucial observation is that as the carrying capacities are made finite, depending on their exact values, any of the composite states can be realized for any combination of the incentives (compare any row of FIG.~\ref{fig3_effect_of_finiteness} with the second row of FIG.~\ref{fig2_comparision}) even if the state is forbidden in the case of infinite carrying capacity. It means that depending on the values of carrying capacities, the TOC can be averted when it is inevitable in an infinite population, or the TOC can be caused when it is avertable in an infinite population. Furthermore, it is crystal clear from the inspection of the trends along any column that the modified incentives are the relevant parameters in the finite carrying capacity case.

\section{Conclusions}
The common-pool resource harvesting is an interesting topic of bio-economical research~\cite{clark1976mathematical} and the eco-evolutionary dynamics that models it has an even wider application: It arguably~\cite{tilman2020nc} models the feedback between plants and soil-microbes~\cite{Bever1997} and also models the decision making~\cite{Rand2017} by agents in the co-evolving environment. It goes without saying that the assumption of an infinite fixed population is quite an unrealistic one in such cases, and hence this paper's contribution in extending the eco-evolutionary dynamics to include the effect of the growing consumer population is very pertinent.

To this end, we have derived the mean-field deterministic eco-evolutionary dynamics from a stochastic birth-death process keeping the finiteness of the populations---both the consumer and the resource---in mind. We have shown that the prevention of the TOC can be effected due to the finiteness of the consumer population through the emergence of bistability and limit cycles in the state space. More importantly, we have systematically investigated the hitherto neglected case where the resource's growth rate lies between the two distinct harvesting rates employed by the consumers. The realization and the prevention of the TOC have been found to be crucially dependent on the incentives. We have witnessed the possibility of both the component and the collapsing TOC. Furthermore, we have found that the collapsing TOC is very much dependent on the initial condition: For a given set of parameters, even if the collapsing TOC appears to be the ultimate fate, it is not so for all possible initial conditions---for the same incentives, sometimes the TOC can be completely bypassed owing to the phenomenon of bistability.

One may tweak the eco-evolutionary dynamics with a view to understanding how different strategies---e.g., reward and punishment~\cite{Mondal2022}---of averting TOC may be effective. Our setup dealing with the finite population may further be explored to study the effect of demographic noise and fluctuating environmental noise~\cite{foster1990tpb,stollmeier2018unfair,Huang2015}. Before we end, we remark that in the light of the very interesting experiments on the interplay between evolution and ecology in changing populations---e.g., in yeast~\cite{sanchez2013feedback} and bacterial~\cite{Wienand2015, Becker2018} populations---we are hopeful that some experiments with microbes~\cite{Korolev2011_1,Korolev2011_2,Frey2010,Pfeiffer2005,Li2015,Lenski2001} may be designed to validate the ideas expounded in this paper.

\acknowledgements
Researches of JDB and DC have been respectively supported by Prime Minister's Research fellowship (govt. of India) and J.C. Bose National fellowship (SERB, India). SC acknowledges support of SERB (DST, govt. of India) project MTR/2021/000119.


\bibliography{Bairagya_etal.bib}
\end{document}